\documentclass{aa}  

\usepackage{graphicx}
\usepackage{cleveref}
\usepackage{amsmath}
\usepackage{txfonts}
\usepackage{natbib}
\bibpunct{(}{)}{;}{a}{}{,} 
\newcommand{\Msun}{{M}_{\odot}}

\begin{document}

   \title{Energetic Explosions from Collisions of Stars at Relativistic Speeds in Galactic Nuclei}


   \author{B. X. Hu
          \inst{1}
          \and
          A. Loeb\inst{2}
          }

   \institute{Department of Physics, Harvard University\\
              \email{bhu@g.harvard.edu}
         \and
             Department of Astronomy, Harvard University\\
             \email{aloeb@cfa.harvard.edu}
             }

   \date{Received Month Day, Year; accepted Month Day, Year}

 
  \abstract
   {}
   {We consider collisions can can occur between stars moving near the speed of light around supermassive black holes (SMBHs) with mass $M_{\bullet}\gtrsim10^8\,\Msun$, without being tidally disrupted. In this approximate SMBH range, for sun-like stars, the tidal-disruption radius is smaller than the SMBH's event horizon; therefore we do not expect to observe tidal disruption events.}
   {Differential collision rates are calculated by defining probability distribution functions for various parameters of interest such as the impact parameter, distance from SMBH at time of collision, relative velocity between the two colliding stars, and the masses of the two colliding stars. The relative velocity parameter is drawn from an appropriate distribution function for SMBHs. We integrate over all parameters to arrive at a total collision rate for a galaxy with a specific SMBH mass. We then consider how the stellar population in the vicinity of the SMBH is depleted and replenished over time, and calculate the effect this can have on the collision rate over time. We further calculate the differential collision rate as a function of total energy released, energy released per unit mass lost, and galactocentric radius.}
   {The overall rate for collisions taking place in the inner $\sim1$ pc of galaxies with $M_{\bullet}=10^8,10^9,10^{10}\,\Msun$ are $\Gamma\sim2.2\times10^{-3},2.2\times10^{-4},4.7\times10^{-5}$ yr$^{-1}$, respectively. The most common collisions will release energies on the order of $\sim10^{49}-10^{51}$ erg, with the energy distribution peaking at higher energies in galaxies with more massive SMBHs. In addition, we show example light curves for collisions with varying parameters, and find that the peak luminosity could reach or even exceed that of superluminous supernovae (SLSNe), although with light curves with much shorter duration.}
   {Weaker events could initially be mistaken for low-luminosity supernovae. In addition, we note that these events will likely create streams of debris that will accrete onto the SMBH and create accretion flares that may resemble tidal disruption events (TDEs).}


   \maketitle
%

\section{Introduction} \label{sec:intro}

Supernova explosions release of order $10^{51}$ ergs of energy, originate from runaway ignition of degenerate white dwarfs \citep{hillebrandt2000} or the collapse of a massive star \citep{woosley1995, barkat1967}. \cite{rubin2011} and \cite{balberg2013} considered a separate, rare kind of explosive event from collisions between hypervelocity stars in galactic nuclei. The cluster of stars builds up over time and reaches a steady state condition in which the rate of stellar collisions is similar to the formation rate of new stars. A simplified model for the explosion light curve with the "radiative zero" approach by Arnett \citep{arnett1996}, which assumes that the shocked material has uniform density and temperature and a homologous velocity profile, shows that the resulting light curve would have an average luminosity on the order of $\sim2\times10^{41}$ erg s$^{-1}$, on par with faint conventional supernovae. Furthermore, the light curve would be expected to include a long flare due to the accretion of stellar material onto the supermassive black hole (SMBH) at the center of the galaxy. \cite{rubin2011} also considered mass loss from collisions between stars at the galactic center in order to constrain the stellar mass function.

In this work, we consider high-speed stars at galactic centers, where there exists a much higher stellar velocity dispersion \citep{sellgren1990}. Approaching stars can be tidally disrupted by a SMBH at the tidal-disruption radius, $r_{\mathrm{T}}\sim R_{\star}(M_{\bullet}/M_{\star})^{1/3}$, with $R_{\star}$ the radius of the star and $M_{\bullet}$ and $M_{\star}$ the masses of the black hole and star, respectively. For sun-like stars, the tidal-disruption radius is smaller than the black hole's event horizon radius $r_{\mathrm{s}}=2GM/c^2$ for black hole masses $\gtrsim10^8\,\Msun$ \citep{stone2019}. For maximally spinning black holes, tidal disruption events (TDEs) can be observed for sun-like stars near SMBHs as large as $\sim7\times10^8\,\Msun$ \citep{kesden2012}. In this work we consider SMBHs with masses $M_{\bullet}\gtrsim10^8\,\Msun$. It is necessary to restrict our study to these rarer, higher-mass SMBHs because we are interested in stars moving at extremely high velocities that could only be achieved very close to the center of the galaxy. At these small radii, in galaxies with lower-mass SMBHs the stars would be tidally disrupted before they could collide. As such, we only consider galaxies in which TDEs would be unlikely to be observed and in which a high-energy stellar collision close to the center of the galaxy could actually occur. We adopt a Newtonian approach and ignore the effects of general relativity near the SMBH because the chances of collisions to occur in a region where they would matter are extremely small.

Surveys from the last two decades such as the Sloan Digital Sky Survey \citep[SDSS,][]{frieman2008}, Palomar Transient Factory \citep[PTF,][]{rau2009}, Zwicky Transient Factory \citep[ZTF,][]{ztf}, Pan-STARRS \citep{scolnic2018}, and others \citep{guillochon2017}, have greatly increased the number of supernovae detected. In addition to detecting many more already well-understood classes of supernovae, previously unheard of transients were also detected, such as superluminous supernovae \citep{galyam2012, bose2018, galyam2019}, rapidly-decaying supernovae \citep{perets2010, kasliwal2010, prentice2018, nakaoka2019, tampo2020}, and transients with slow temporal evolution \citep{taddia2016, arcavi2017, dong2020, gutierrez2020}. These discoveries have challenged existing theories of transients and suggest that a much broader range of events remain to be detected. The Vera C. Rubin observatory is expected to start operation in 2025 and to detect hundreds of thousands of supernovae a year over a ten-year survey \citep{lsst2019}. 

The outline of this paper is as follows. In section \ref{sec:method}, we describe how we simulate stellar collisions and calculate light curves. In section \ref{sec:results}, we provide the results of our calculations. In section \ref{sec:observedrates}, we estimate the observed rates of our events. Finally, in section \ref{sec:discussion} we summarize our main conclusions. 

\section{Method} \label{sec:method}

\subsection{Explosion Parameters} \label{ssec:explosion}

\cite{rubin2011} provide the differential collision rate between two species of stars, labeled "1" and "2", at some impact parameter $b$ with distribution functions $f_1$ and $f_2$ and velocities $\vec{v}_1$ and $\vec{v}_2$,
\begin{equation}\label{eq:1}
\begin{split}
\ d\Gamma=&f_1\left(r_{\mathrm{gal}},\vec{v}_1\right)d^3v_1f_2\left(r_{\mathrm{gal}},\vec{v}_2\right)d^3v_2\times \\
    &|\vec{v}_1-\vec{v}_2|2\pi b\,db\,4\pi r^2_{\mathrm{gal}}dr_{\mathrm{gal}},
\end{split}
\end{equation}
assuming spherical symmetry, with dependence only on galocentric radius, $r_{\mathrm{gal}}$. 

Adopting a power-law present-day mass function (PDMF), $\xi\equiv dn/dM\propto M^{-\alpha}$, Eq. \eqref{eq:1} simplifies to
\begin{equation}\label{eq:2}
\begin{split}
\ d\Gamma=&8\pi^2v_{\mathrm{rel}}f(r_{\mathrm{gal}},v_{\mathrm{rel}})K^2\left(r_{\mathrm{gal}}\right)M_1^{-\alpha}M_2^{-\alpha}r_{\mathrm{gal}}^2b\times \\
    &db\,dr_{\mathrm{gal}}\,dv_{\mathrm{rel}}\,dM_1\,dM_2.
\end{split}
\end{equation}
The relative velocity between the stars is $v_{\mathrm{rel}}=|\vec{v}_1-\vec{v}_2|$, and $K\left(r_{\mathrm{gal}}\right)$ is a normalization constant which can be solved for from the density profile,
\begin{equation}\label{eq:3}
\ \rho\left(r_{\mathrm{gal}}\right)=K\left(r_{\mathrm{gal}}\right)\left(\frac{M_{\mathrm{max}}^{2-\alpha}-M_{\mathrm{min}}^{2-\alpha}}{2-\alpha}\right).
\end{equation}
The stellar density profile is adapted from \cite{tremaine1994},
\begin{equation}\label{eq:4}
\ \rho_{\eta}(r_{\mathrm{gal}})\equiv\frac{\eta}{4\pi}\frac{r_sM_{\mathrm{\star}}}{r_{\mathrm{gal}}^{3-\eta}(r_s+r_{\mathrm{gal}})^{1+\eta}},
\end{equation}
where we adopt the commonly-used index $\eta=2$ \citep{hernquist1990}, $M_{\star}$ is the total mass of the host spheroid, and $r_s$ is a distinctive scaling radius. We use the following relation between the mass of the black hole $M_{\bullet}$ and the mass of the host spheroid $M_{*}$ \citep{graham2012},
\begin{equation}\label{eq:5}
\ \log\left(\frac{M_{\bullet}}{\Msun}\right)=\alpha+\beta\log\left(\frac{M_{\star}}{7\times10^{10}\,\Msun}\right),
\end{equation}
with best-fit values $\alpha=8.4$ and $\beta=1.01$. For $M_{\bullet}=10^8,10^9,10^{10}\,\Msun$, we find spheroid masses of $M_{\star}\sim2.8\times10^{10},2.7\times10^{11},2.7\times10^{12}\,\Msun$, respectively. Using our chosen parameters and the data from \cite{sahu2020}, we take the scaling radius as $r_s\sim0.8,6,50$ kpc, respectively. 

Based on Eq. \eqref{eq:2}, we define probability distribution functions (PDFs) for the parameters $b$, $r_{\mathrm{gal}}$, $v_{\mathrm{rel}}$, $M_1$, and $M_2$. We assume a Salpeter-like mass function and take $\alpha=2.35, M_{\mathrm{min}}=0.1\,\Msun,$ and $M_{\mathrm{max}}=125\,\Msun$. For the impact parameter $b$, we take $dP/db\propto b$, where we take $b_{\mathrm{min}}=0$ and $b_{\mathrm{max}}=R_1+R_2$, the sum of the radii of the colliding stars. This in turn requires the values of the two two radii $R_1$ and $R_2$. We use the stellar $M-R$ relation,
\begin{equation}\label{eq:9}
\ \log R=a+b\log M,
\end{equation}
with $a=0.026$ and $b=0.945$ for $M<1.66\,\Msun$, and $a=0.124$ and $b=0.555$ for $M>1.66\,\Msun$ \citep{demircan1991}. The PDF for the galactocentric radius $r_{\mathrm{gal}}$ can be calculated from the density profile, $dP/dr_{\mathrm{gal}}\propto\rho\left(r_{\mathrm{gal}}\right)r_{\mathrm{gal}}^2$, where we take $r_{\mathrm{min}}=10^{-5}$ pc and $r_{\mathrm{max}}=200$ pc. However, the relevant range of interest for this work (i.e. where high-velocity collisions are most likely to take place) is actually only from roughly $r_{\mathrm{min}}$ to ~1 pc, the latter distance which we call $r_{\mathrm{cap}}$. 

\cite{tremaine1994} provides the distribution function $f_{1,2}$,
\begin{equation}\label{eq:7}
\begin{split}
\ f\left(r_{\mathrm{gal}},\vec{v}_1\right)&=\left(\frac{\Gamma(2)}{2^{3/2}\pi^{5/2}M_{\bullet}\Gamma(1/2)}\right)\left(\frac{M_*^2}{a^3v_g^3}\right)\left(\frac{-a}{GM_*}\right)^{-1/4}\times \\
    &\left(-\frac{GM_*}{a+r_{\mathrm{gal}}}-\frac{GM_{\bullet}}{r_{\mathrm{gal}}}+\frac{v_1^2}{2}\right)^{-1/4},
\end{split}
\end{equation} 
with $\Gamma(x)$ being the gamma function, $M_{\bullet}$ the mass of the supermassive black hole (SMBH), $M_*$ the stellar mass of the galaxy, and $a$ and $v_g$ constants defined by the galaxy being studied. We were unable to find an analytical expression for the relative velocity distribution corresponding to this individual velocity distribution, but were able to solve for it numerically by sampling pairs of velocities from Eq. \eqref{eq:2} and calculating $v_{\mathrm{rel}}$ as the magnitude of the difference between the two vectors (with the angles corresponding to the vectors being randomly assigned). We found that resulting histogram of $v_{\mathrm{rel}}$ could be well-approximated by a Maxwellian, with the parameters determined by a fitting function. We use this best-fit Maxwellian function as $f\left(r_{\mathrm{gal}},\vec{v}_{\mathrm{rel}}\right)$ in our calculations instead of instead of the numerically-derived interpolation because we find that our calculations can run significantly faster when using analytic expressions. We note that the Maxwellian expression actually provides a slightly more conservative estimate of our final results compared to the numerically-derived expression because it both peaks at a lower velocity for a given radius and has a less-prominent high-velocity tail.

To run a Monte Carlo integration, we draw a fixed number $N=10^6$ of sample values from each of the probability distributions. Each sample is meant to represent two stars with known masses ($M_1$, $M_2$) and radii ($R_1$, $R_2$) colliding with some know relative velocity $v_{\mathrm{rel}}$ and impact parameter $b$ at some galactocentric radius $r_{\mathrm{gal}}$. We use a Monte Carlo estimator to calculate the multidimensional integral,
\begin{equation}\label{eq:12}
\ \Gamma=\int d\Gamma(b, r_{\mathrm{gal}}, v_{\mathrm{rel}}, M_1, M_2)db\,dr_{\mathrm{gal}}\,dv_{\mathrm{rel}}\,dM_1\,dM_2.
\end{equation}
For a given collision, the kinetic energy of the ejecta is estimated from collision kinematics as,
\begin{equation}\label{eq:13}
\ E_{\mathrm{ej}}=\mu v_{\mathrm{rel}}^2\frac{A_{\mathrm{int}}(R_1,R_2,b)}{\pi\mathrm{Min}(R_1,R_2)^2}=\frac{M_1M_2v_{\mathrm{rel}}^2}{2(M_1+M_2)}\frac{A_{\mathrm{int}}(R_1,R_2,b)}{\pi\mathrm{Min}(R_1,R_2)^2},.
\end{equation}
with $\mu$ the reduced mass and $A_{\mathrm{int}}(R_1,R_2,b)$ the area of intersection of the collision,

\begin{equation}\label{eq:14}
\begin{split}
\ A_{\mathrm{int}}&(R_1,R_2,b)=R_1^2\cos^{-1}\left(\frac{b^2+R_1^2-R_2^2}{2bR_1}\right) \\
        &+R_2^2\cos^{-1}\left(\frac{b^2+R_2^2-R_1^2}{2bR_2}\right) \\
	&-\frac{1}{2}\sqrt{(-b+R_1+R_2)(b+R_1-R_2)(b-R_1+R_2)(b+R_1+R_2)}
\end{split}
\end{equation}

We define the enclosed stellar mass $\mathcal{M}(r_{\mathrm{gal}})\equiv\int_{r_{\mathrm{min}}}^{r_{\mathrm{gal}}}\rho(r_{\mathrm{gal}}')4\pi r_{\mathrm{gal}}'^2dr_{\mathrm{gal}}'$. We can roughly calculate the mass lost in a collision between two stars as,
\begin{equation}\label{eq:15}
\ M_{\mathrm{lost}}\approx M_1\left(\frac{V_{\mathrm{int}}(R_1,R_2,b)}{\frac{4\pi}{3}R_1^3}\right)+M_2\left(\frac{V_{\mathrm{int}}(R_1,R_2,b)}{\frac{4\pi}{3}R_2^3}\right).
\end{equation}
with $V_{\mathrm{int}}(R_1,R_2,b)$ the volume of intersection between the two spherical stars for some impact parameters $b$, 
\begin{equation}\label{eq:15}
\begin{split}
\ V_{\mathrm{int}}&(R_1,R_2,b)=\frac{\pi(R_1+R_2-b)^2}{12b}\times \\
    &\frac{(b^2+2bR_2-3R_2^2+2bR_1+6R_1R_2-3R_1^2)}{12b}.
\end{split}
\end{equation}

We note that this is, at best, an order-of-magnitude approximation, and that the volume fractions in Eq. \eqref{eq:15} roughly approximate the fraction of the relative kinetic energy deposited in each star via the thermalization of shockwaves. We use this to calculate $M_{\mathrm{lost,avg}}\sim0.08\,\Msun$ as the average mass lost in a collision.

We note that from Eqs. \eqref{eq:2} and \eqref{eq:3}, $\Gamma\propto\rho^2$. We can use this relation, along with our initially assumed stellar density profiles, resulting depletion rates, and some assumed star formation rate, to calculate a basic dynamic stellar density profile that reflects stars both forming and being destroyed through collisions. For this calculation, we divide the area of interest into equal logarithmic bins in $r_{\mathrm{gal}}$. In each bin, we start with the stellar density as given by Tremaine et al. 1994, as well as the corresponding collision rate Gamma for that bin. We assume a static star formation rate of approximately $1, 5$, and $10\,M_{\odot}/\mathrm{yr}$ for $10^8, 10^9,$ and $10^{10}\,M_{\odot}$ black hole galaxies \citep{behroozi2019}, respectively. We further assume that this star formation rate is uniformly distributed over the area enclosed by the galaxy's bulge effective radius $R_e$, which we calculate from the following $M_{\bullet}-R_e$ relation \citep{sani2011},
\begin{equation}\label{eq:15.5}
\ \log M_{\bullet}=(8.22\pm0.08)+(0.88\pm0.17)\times\left[\log\left(R_e/\mathrm{kpc}\right)-0.4\right].
\end{equation}
To estimate how the stellar density will change over time, we choose a relatively small time step, such as 100 years, and then subtract the number of stars lost to collisions and add the number of stars formed. We can then calculate a new value of $\rho$ in this radius bin, and from this a new value of $\Gamma$. This process is repeated as needed until a set amount of time has passed.

\subsection{Light Curves} \label{ssec:lightcurves}

To calculate light curves for star-star collisions, we follow the analytic modeling approach of Arnett \citep{arnett1980, arnett1982}. This approach assumes that the ejecta is expanding homologously, radiation pressure dominates over  gas pressure, the luminosity can be described by the spherical diffusion equation, and that the ejecta is characterized by a constant opacity \citep{khatami2019}. Given these assumptions, the light curve is described by,
\begin{equation}\label{eq:18}
\ L(t)=\frac{2}{\tau_d^2}e^{-t^2/\tau_d^2}\int_0^t t'L_{\mathrm{heat}}(t')e^{t^{'2}/\tau_d^2}dt',
\end{equation}
where $\tau_d$ is the characteristic diffusion time,
\begin{equation}\label{eq:19}
\ \tau_d=\left[\frac{3}{4\pi}\frac{\kappa M_{\mathrm{ej}}}{v_{\mathrm{ej}}c}\frac{1}{\xi}\right]^{1/2},
\end{equation}
with $M_{\mathrm{ej}}$ and $v_{\mathrm{ej}}$ the mass and velocity of the ejecta, respectively, $\kappa$ taken to be the electron scattering opacity $\kappa_{\mathrm{es}}=0.2(1+X)$ cm$^2$/g, where $X$ is the fractional abundance of hydrogen, and $\xi=\pi^2/3$ \citep{khatami2019}. $L_{\mathrm{heat}}(t')$ is the total input heating rate, which we take to be $L_{\mathrm{int}}(t)=L_0e^{-t/t_s}$, normalized so that for a given collision $\int_0^{\infty}L_{\mathrm{int}}(t)dt=E_{\mathrm{ej}}=\chi M_{\mathrm{ej}}v_{\mathrm{ej}}^2/2$, where $\chi$ is an efficiency factor between 0 and 1. We estimate $t_s$ as $R/t_{\mathrm{ej}}$, where $R$ is the combined diameter of the sum of the two stars' diameters minus the impact parameter, and given $R$ we can calculate $t_{\mathrm{ej}}$ from the velocity of the ejecta (calculated from the kinetic energy of the explosion). The heat, in our case, is supplied very early by the kinetic energy deposited in the collision. We expect that the heating rate rapidly decays, and this is reflected by a very short $t_s$. Given this, we find that the diffusion time given in Eq. (\ref{eq:19}) varies as a factor which we define as $\lambda$,
\begin{equation}\label{eq:20}
\ \tau_d\propto\frac{\kappa^2\chi M_{\mathrm{ej}}^3}{E_{\mathrm{ej}}}\equiv\lambda.
\end{equation}
In our fiducial model, we take $\kappa=0.4$ cm$^2$/g, $\chi=0.5$, $M_{\mathrm{ej}}=1\,\Msun$, and $E_{\mathrm{ej}}=10^{51}$ ergs, and label $\lambda$ with these chosen values $\lambda_0$. We note that we expect there to be an initial shock breakout which should result in a bright flash at very early times \citep{colgate1974, matzner1999, nakar2010}, but that feature is not included in our simplified model. The analytical solutions presented by Arnett were derived in the context of supernovae (SNe). The flux from a SNe can be emitted in multiple wavelength regimes depending on the radioactive decays of newly synthesized isotopes. For type 1A SNe, $\sim85\%$ of the luminosity is emitted at optical wavelengths due to the radioactive decay of nickel-56 \citep{jha2019}. In the case of stellar collisions, more detailed simulations would likely have to be conducted to precisely estimate nickel masses \citep{konyves2020}, if the heating source is more central or evenly mixed \citep{khatami2019}, $\gamma$-ray escape, and other factors that can contribute to emission in various wavelength regimes.

\section{Results} \label{sec:results}

Using the Monte Carlo estimator method described above with $N=10^5$ samples, we estimate total collision rates of $\Gamma=2.2\times10^{-3},2.2\times10^{-4},4.7\times10^{-5}$ yr$^{-1}$ collisions per year for $M_{\bullet}=10^8,10^9,10^{10}\,\Msun$, respectively, in our range of interest, $r_{\mathrm{gal}}<1$ pc. In general, although the spheroid mass is larger for galaxies with more massive SMBHs, the stellar density is overall lower, which results in a lower collision rate. 

\begin{figure}
   \centering
   \includegraphics[width=0.9\hsize]{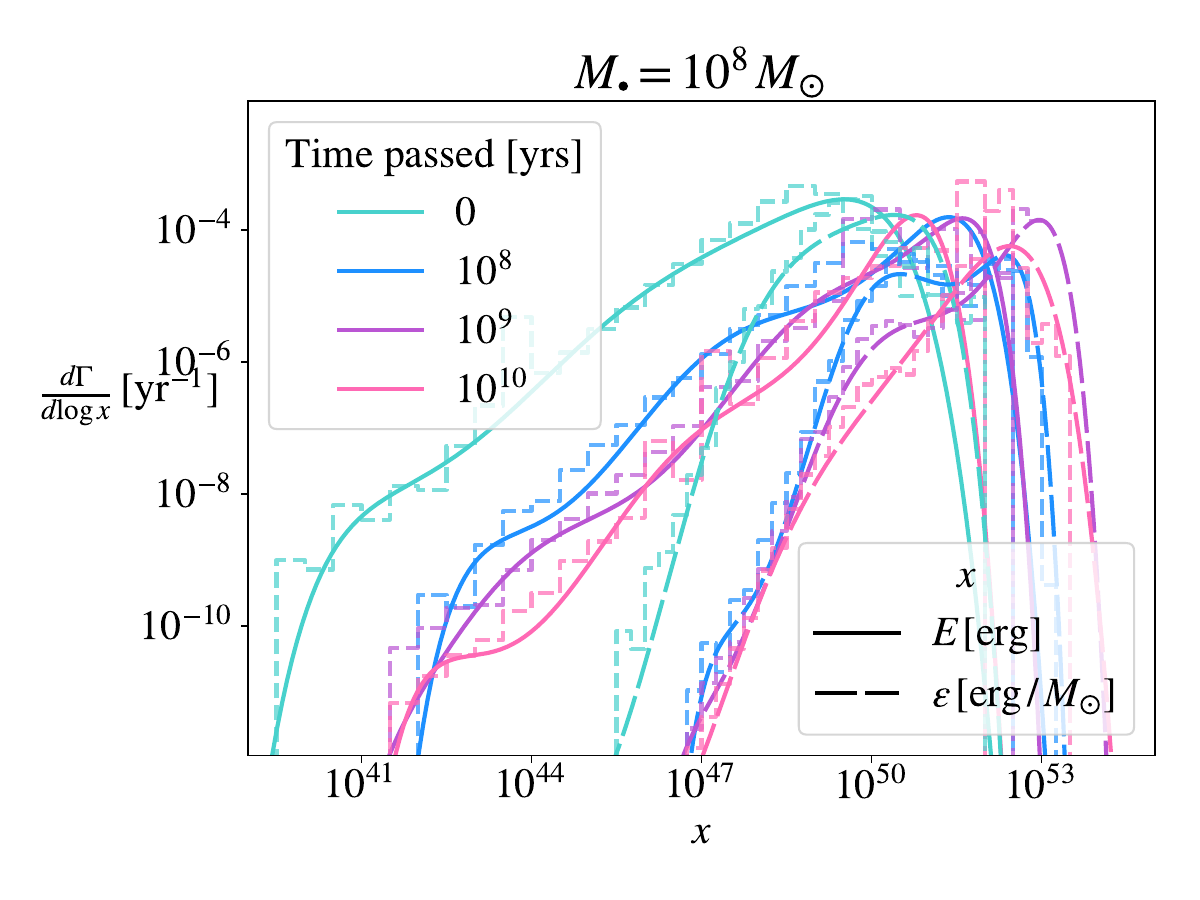}{(a)}
   \includegraphics[width=0.9\hsize]{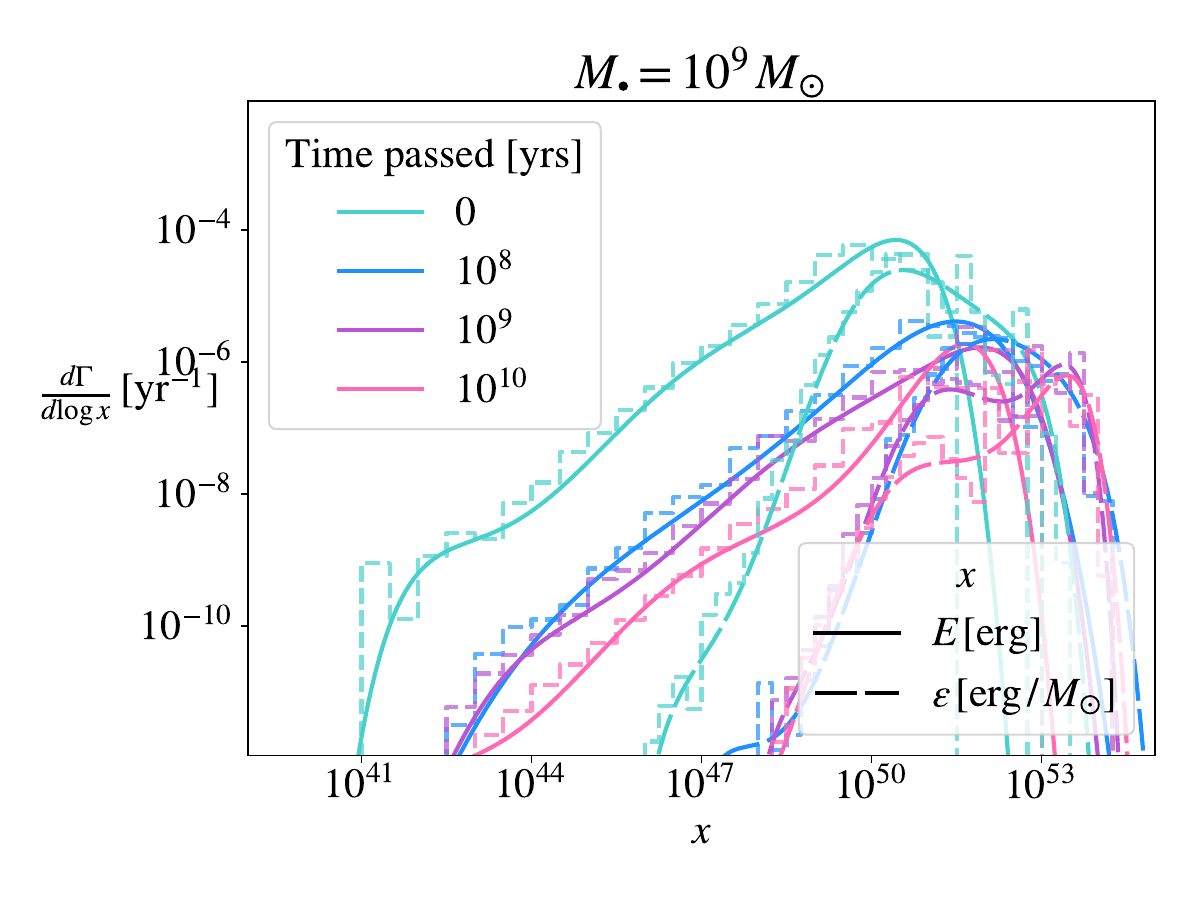}{(b)}
   \includegraphics[width=0.9\hsize]{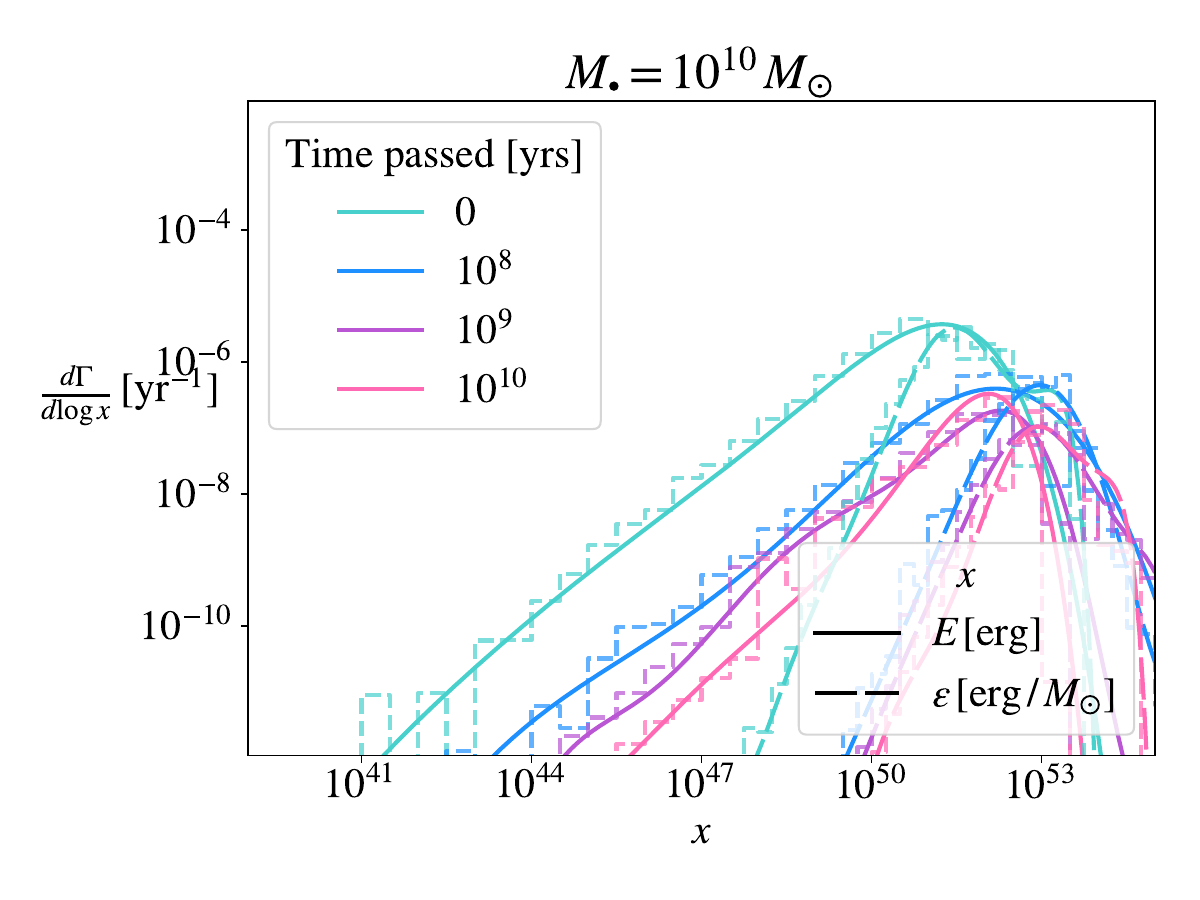}{(c)}
      \caption{The differential collision rate, $d\Gamma$, binned by both logarithmic energy of the ejecta, $E_{\mathrm{ej}}$, and energy of the ejecta per unit mass lost, $\epsilon\equiv E_{\mathrm{ej}}/M_{\mathrm{lost}}$, for (a) $M_{\bullet}=10^8\,\Msun$, (b) $M_{\bullet}=10^{9}\,\Msun$, and (c) $M_{\bullet}=10^{10}\,\Msun$, each for $0,10^8,10^9$, and $10^{10}$ years passed.
              }
         \label{fig:1}
\end{figure}

Figure \ref{fig:1} plots the differential collision rate binned by both logarithmic energy of the ejecta $E_{\mathrm{ej}}$ and energy per unit mass, $\epsilon\equiv E_{\mathrm{ej}}/M_{\mathrm{lost}}$. For the sake of comparison, we estimate the rate of core-collapse supernovae (CCSNe) in similar galaxies as the overall CCSNe volumetric rate \citep{frohmaier2021} normalized by the total star formation rate (SFR) and multiplied by the SFR of galaxies with SMBHs of the same mass \citep{behroozi2019}. Using this prescription, we calculate CCSNe rates of $\Gamma_{\mathrm{CCSNe}}\sim0.01, 0.05, 0.1$, for $M_{\bullet}=10^8,10^9,10^{10}$, respectively. A typical CCSNe ejecta mass is on the order of $M_{\mathrm{lost}}\sim10\,\Msun$ \citep{smartt2009}. These rates are much higher than the collision rates we calculate. However, we note that these CCSNe rates are calculated for entire galaxies, while we only consider the innermost $\sim1$ pc for stellar collisions, so the CCSNe rates we quote should be considered over-estimates for direct comparison purposes. 

We note that although rates have been estimated, we do not compare collision events to superluminous supernovae (SLSNe) because they seem to show preference for low-mass (low-metallicity) environments \citep{leloudas2015, angus2016}.

\begin{figure}
   \centering
   \includegraphics[width=0.9\hsize]{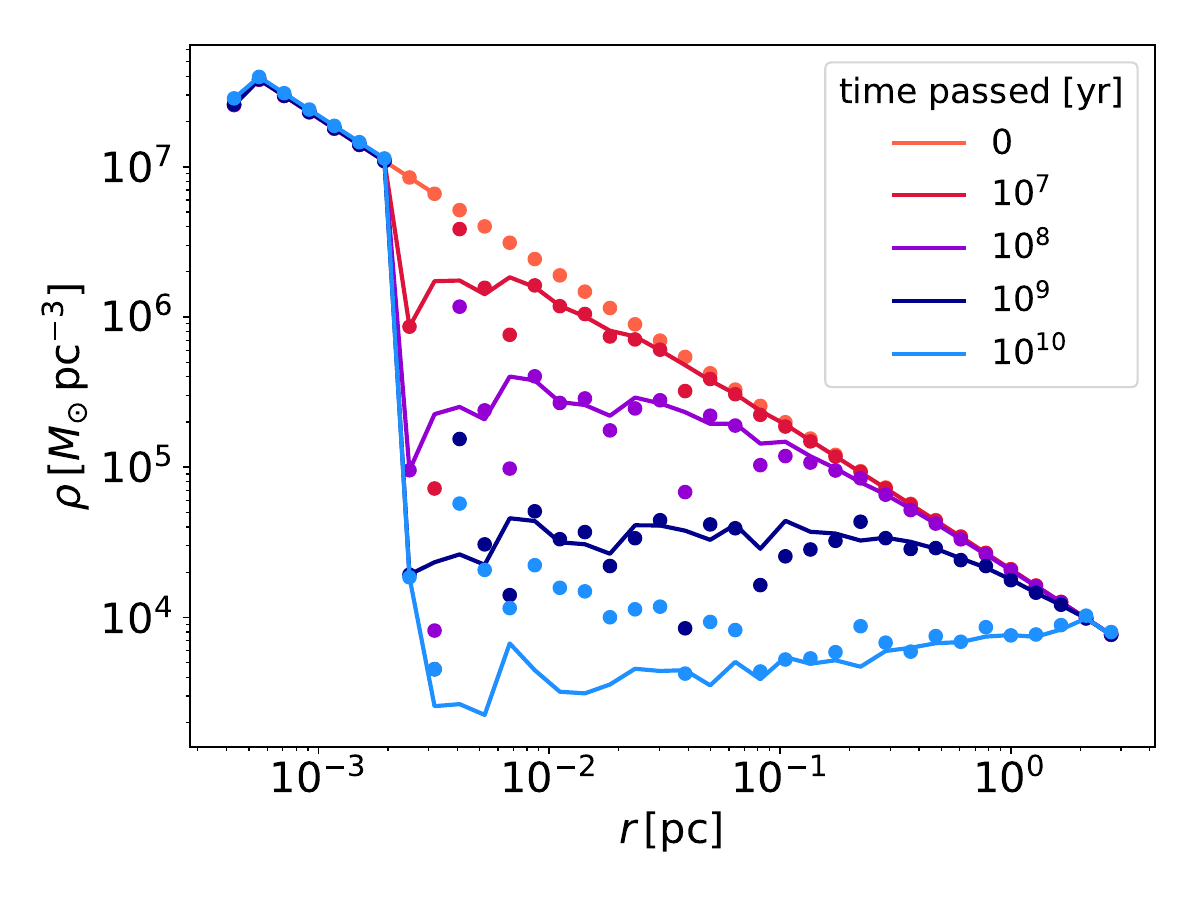}{(a)}
   \includegraphics[width=0.9\hsize]{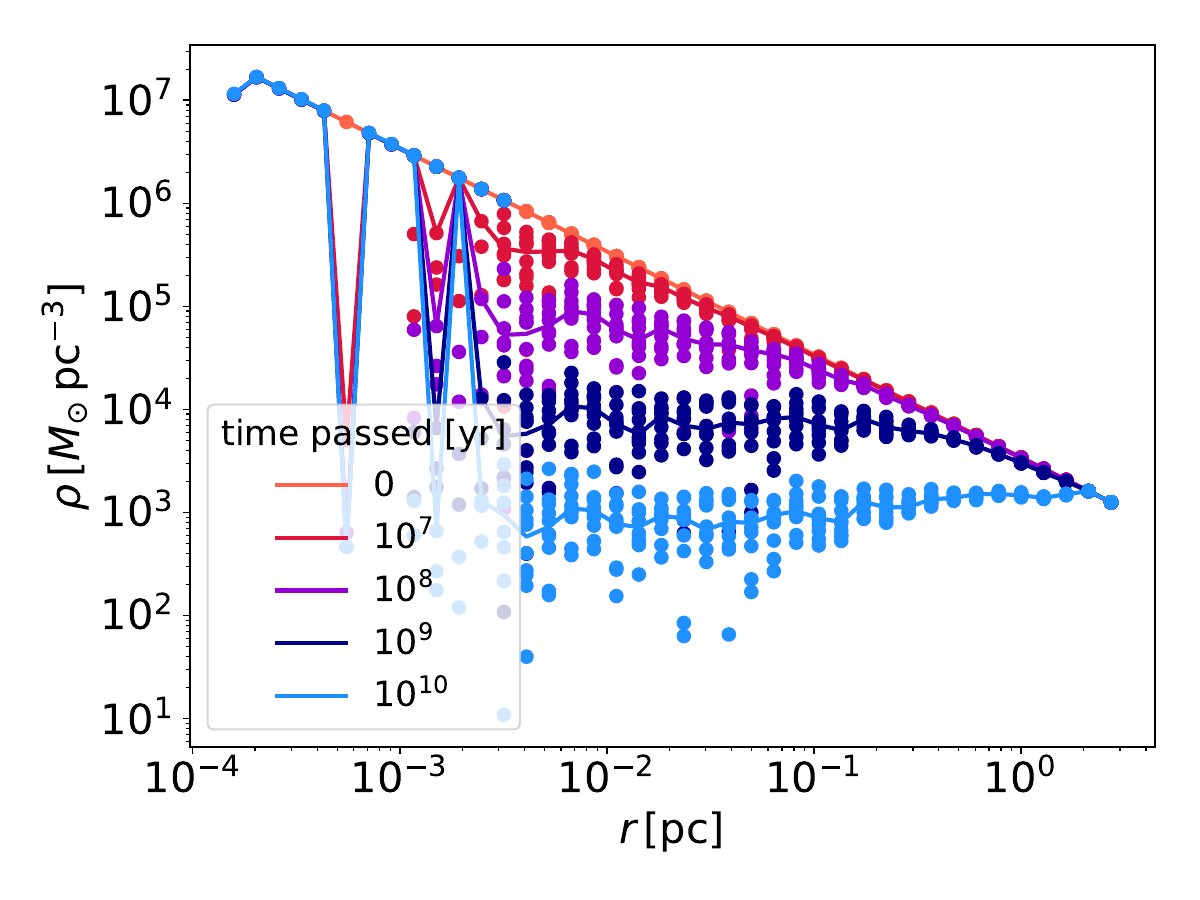}{(b)}
   \includegraphics[width=0.9\hsize]{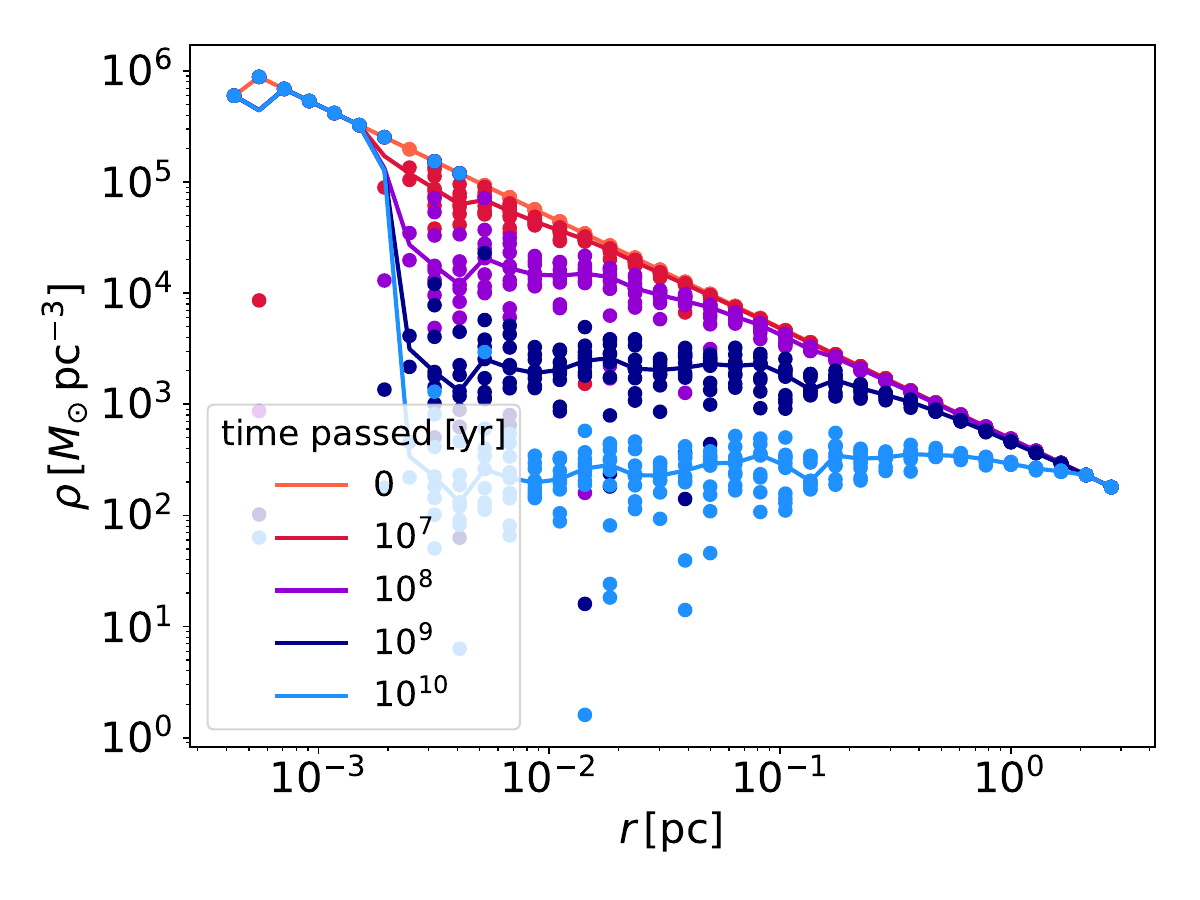}{(c)}
      \caption{Stellar density profiles with $0,10^7,10^8,10^9$, and $10^{10}$ years passed for (a) $M_{\bullet}=10^8\,\Msun$, (b) $M_{\bullet}=10^9\,\Msun$, and (c) $M_{\bullet}=10^{10}\,\Msun$. The purpose of these profiles is to provide a rough idea of how stellar density profiles can change over time as the depletion of stars from collisions is balanced by the continuous formation of new stars.
              }
         \label{fig:2}
\end{figure}

Figure \ref{fig:2} shows stellar density profiles for our galaxy with $0,10^7,10^8,10^9$, and $10^{10}$ years passed. These profiles are calculated from the differential collision rate as a function of galactocentric radius using the profile specified in Eq. \eqref{eq:4}, which we then move forward in time while adding stars from star formation and subtracting them from stellar collisions. 

\begin{figure}
   \centering
   \includegraphics[width=0.9\hsize]{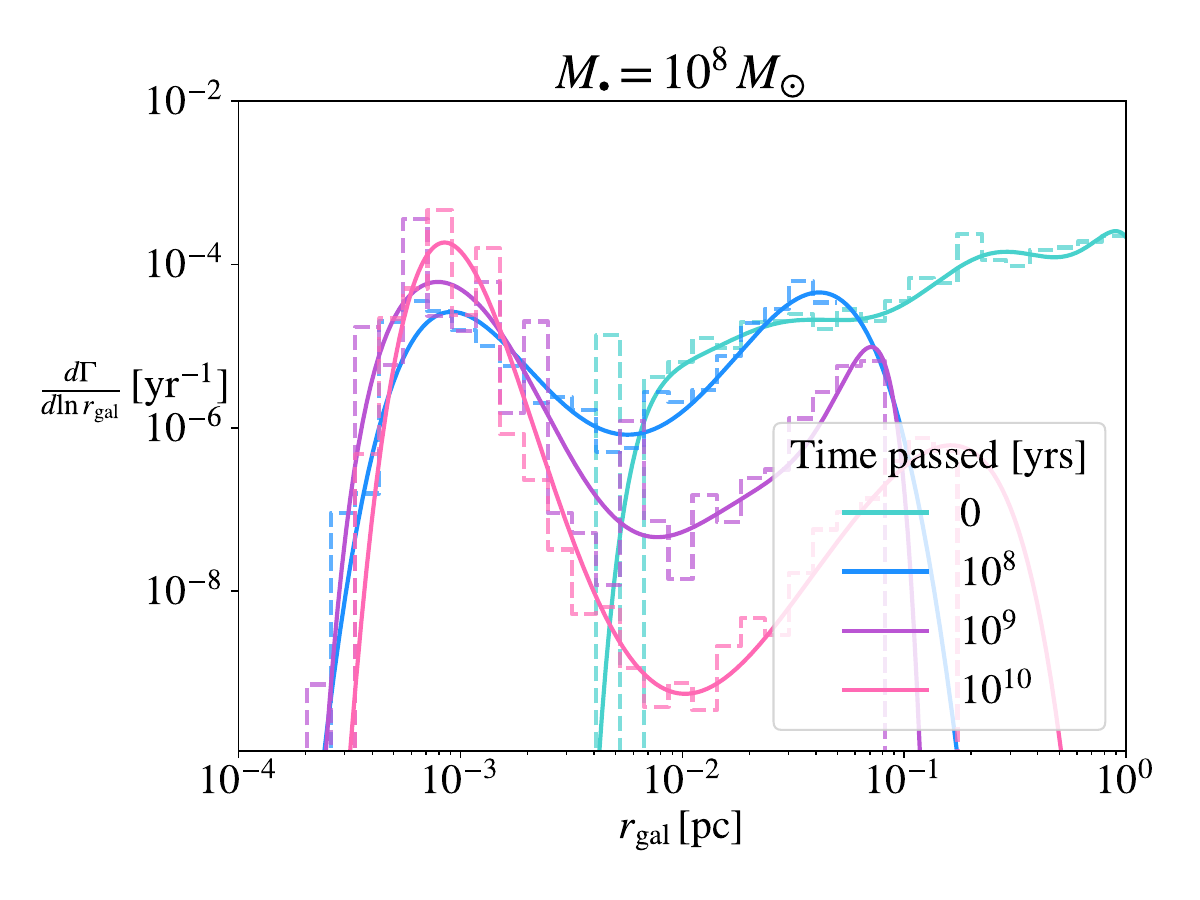}{(a)}
   \includegraphics[width=0.9\hsize]{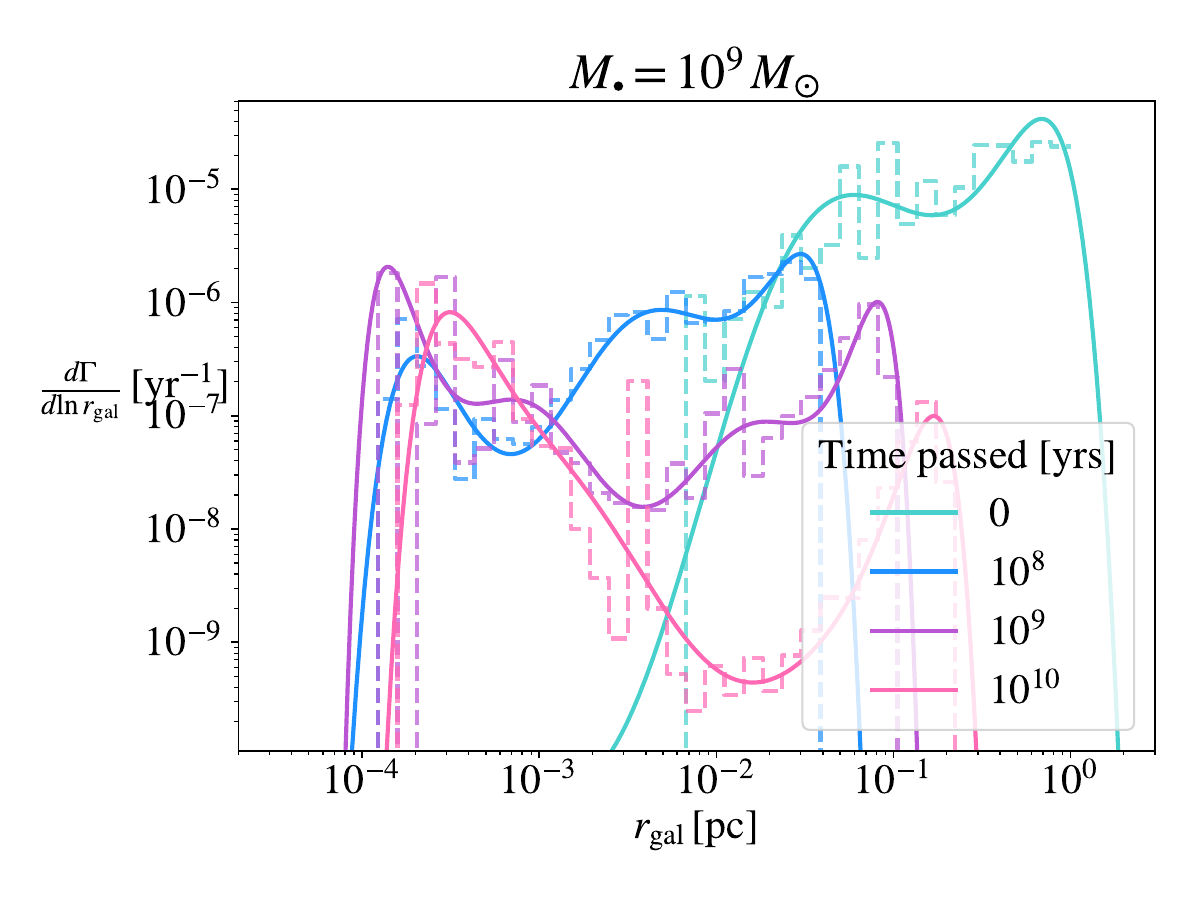}{(b)}
   \includegraphics[width=0.9\hsize]{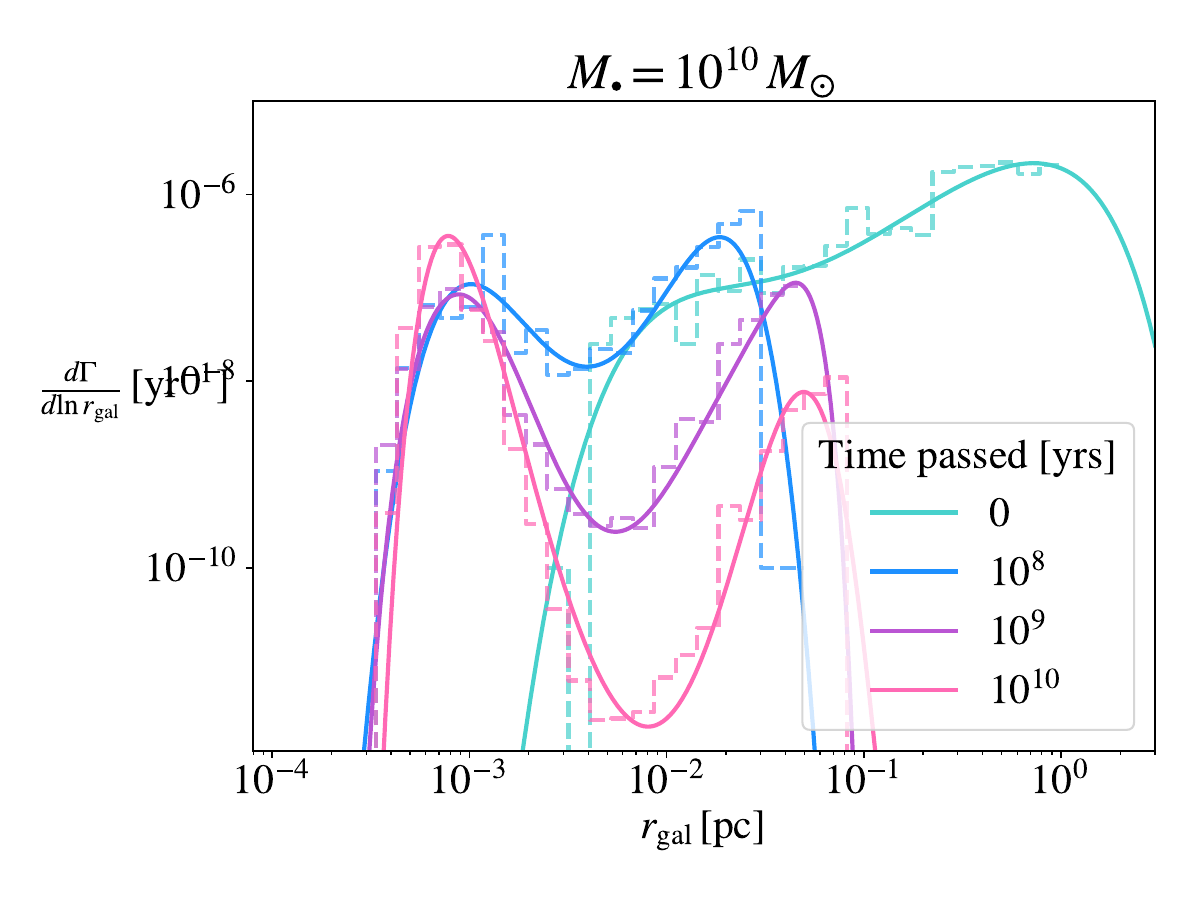}{(c)}
      \caption{The differential collision rate, $d\Gamma$, binned by logarithmic galactocentric radius, $r_{\mathrm{gal}}$, using the stellar density profiles shown in Fig. \ref{fig:2} for (a) $M_{\bullet}=10^8\,\Msun$, (b) $M_{\bullet}=10^9\,\Msun$, and (c) $M_{\bullet}=10^{10}\,\Msun$. From eqs. (\ref{eq:3}-\ref{eq:4}), we have that $d\Gamma\propto\rho^2$ with all other variables fixed.
              }
         \label{fig:3}
\end{figure}

Figure \ref{fig:3} shows the resulting differential collision rate per logarithmic galactocentric radius, $d\Gamma/d\ln r_{\mathrm{gal}}$, with $0,10^8,10^9$, and $10^{10}$ years passed. We use the stellar density profiles calculated and shown in Fig. \ref{fig:2}. We note that although $d\Gamma\propto\rho^2$ with all other variables fixed, for a given stellar density profile the collision rate tends to decrease towards the center of the galaxy, which is expected due to overall smaller decrease in enclosed volume for smaller $r_{\mathrm{gal}}$ since the central density profiles are shallower than $r_{\mathrm{gal}}^{-3/2}$ as a result of their depletion. This is reflected in the $r_{\mathrm{gal}}^2$ term in Eq. \eqref{eq:2}.

\begin{figure}
   \centering
   \includegraphics[width=0.9\hsize]{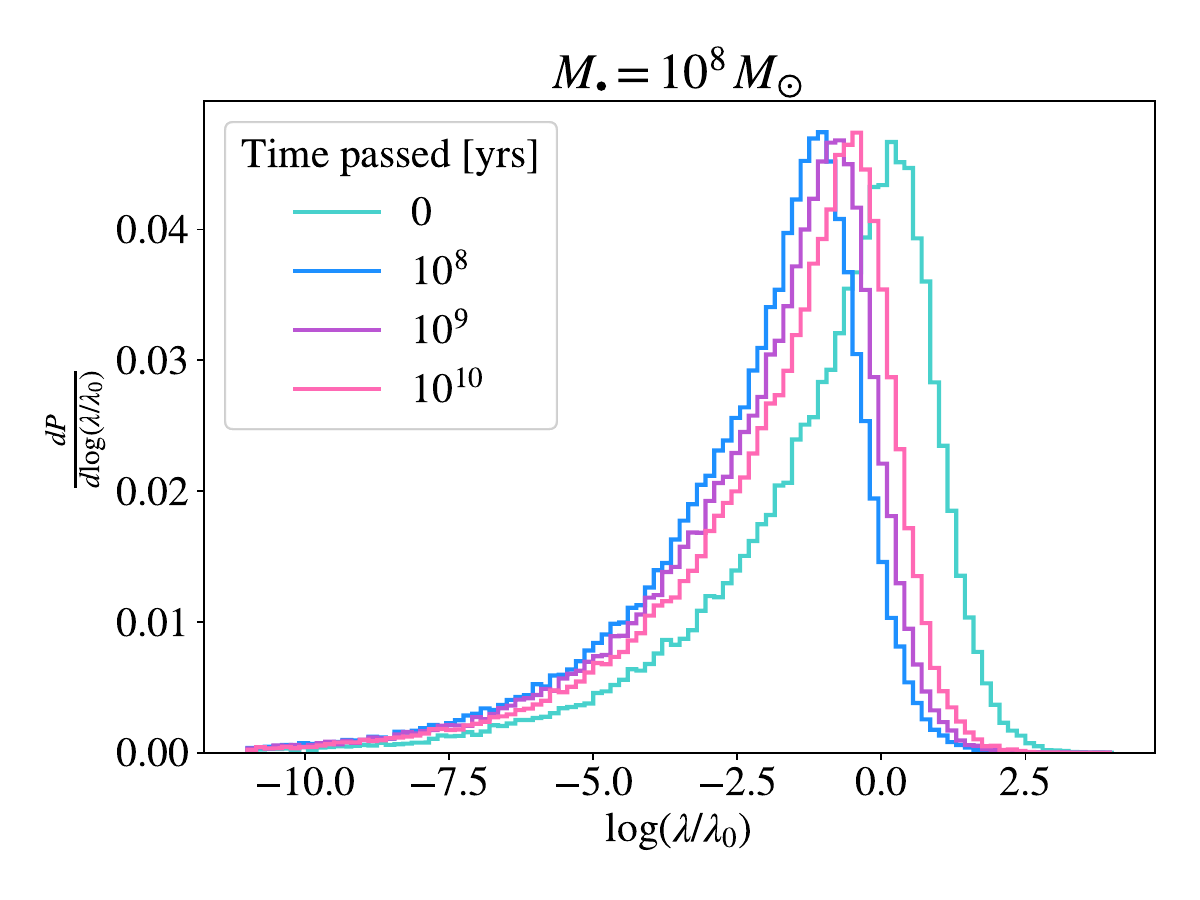}{(a)}
   \includegraphics[width=0.9\hsize]{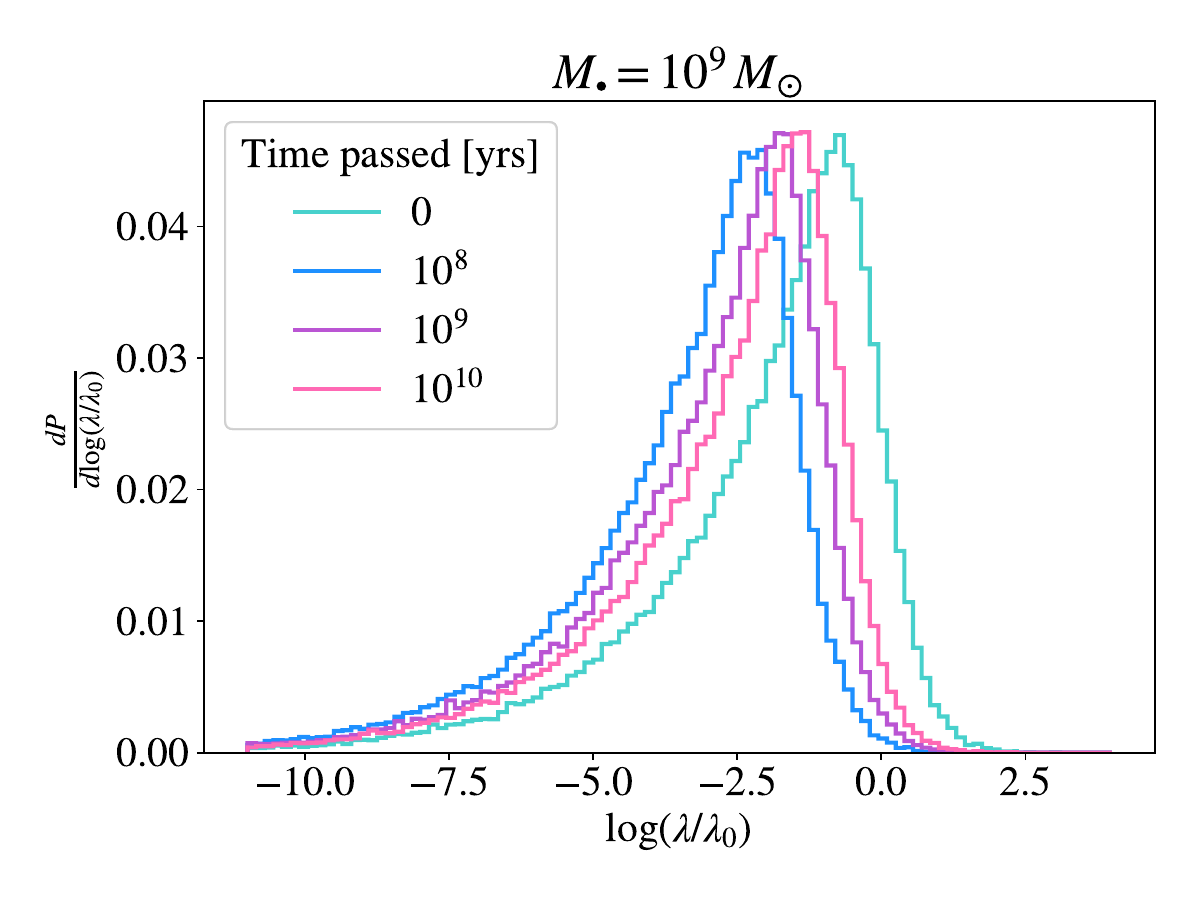}{(b)}
   \includegraphics[width=0.9\hsize]{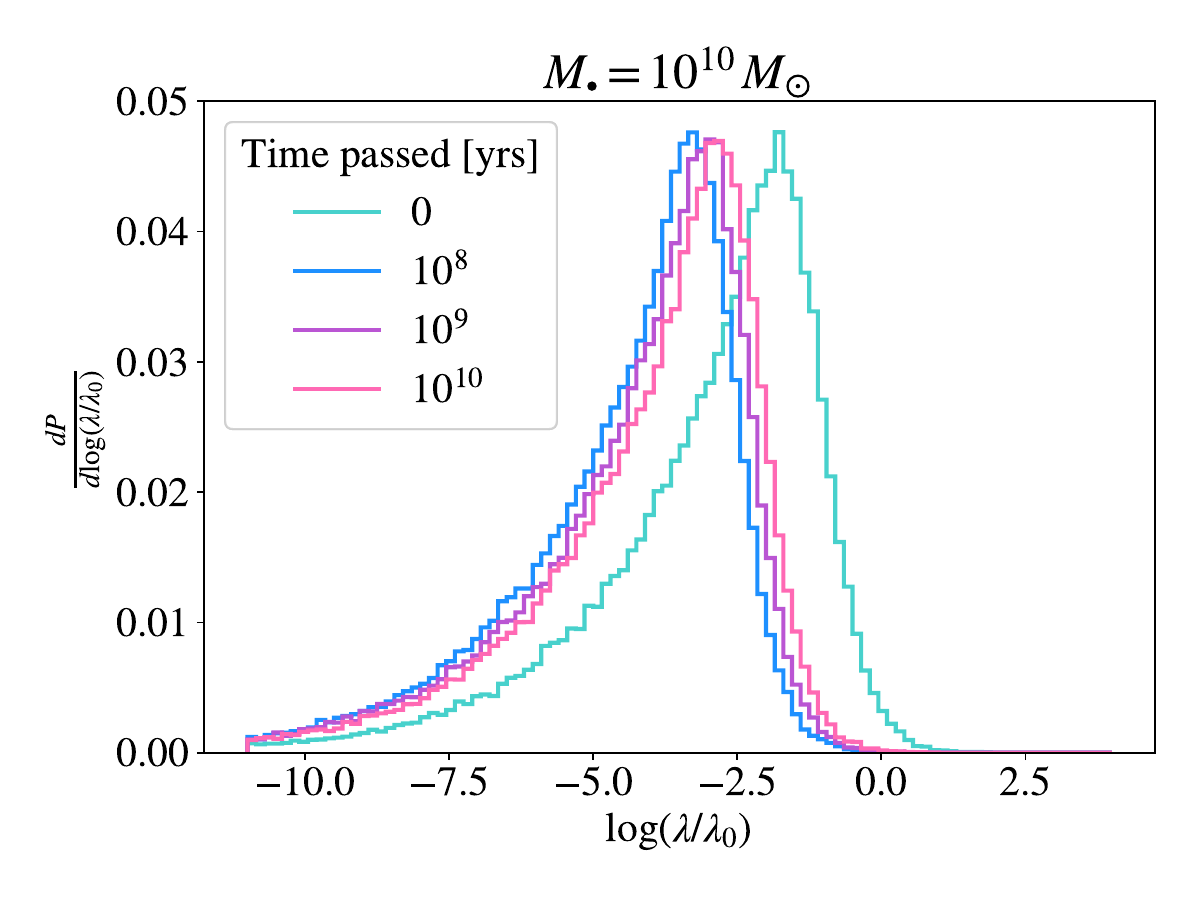}{(c)}
      \caption{The distribution of the variable $\lambda$ with respect to $\lambda_0$ with $0,10^8,10^9$, and $10^{10}$ years passed for (a) $M_{\bullet}=10^8\,\Msun$, (b) $M_{\bullet}=10^9\,\Msun$, and (c) $M_{\bullet}=10^{10}\,\Msun$ The majority of samples fall in the range $\lambda/\lambda_0<1$. The long tail towards lower values of $\lambda$ reflects the high number of grazing impacts, which in turn have very low $M_{\mathrm{ej}}$.
              }
         \label{fig:4}
\end{figure}

\begin{figure}
   \centering
   \includegraphics[width=0.9\hsize]{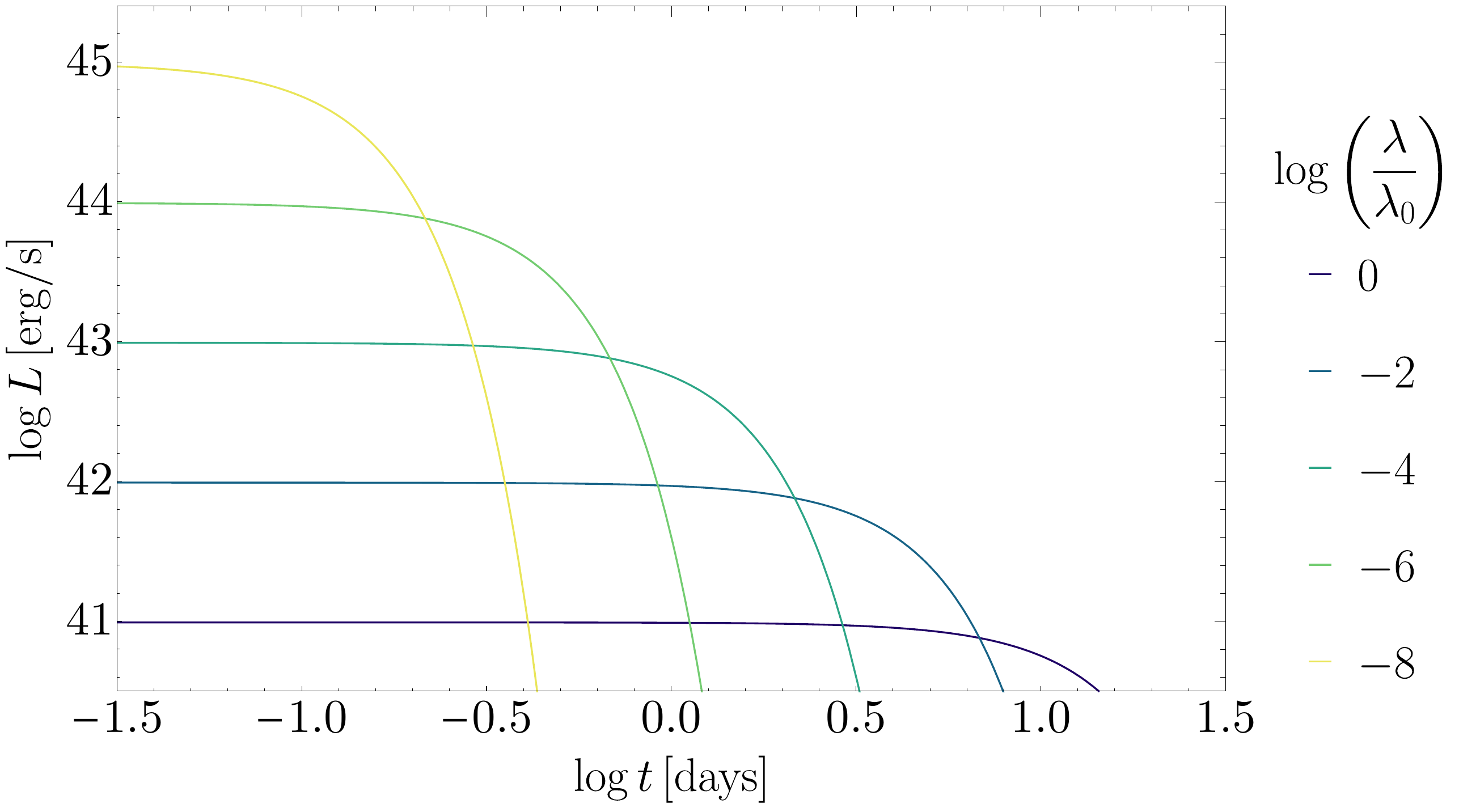}
      \caption{Example light curves based on the distribution of $\lambda/\lambda_0$ using the analytic methods described by \cite{arnett1980, arnett1982}. Although it appears possible for the peak luminosity of a stellar collision to reach or even surpass that of a SLSNe, the extremely short duration makes it less likely that such an event could be detected. However, less luminous events which could initially be mistaken as low-luminosity supernovae could potentially be detected, especially with advances in survey technology.
              }
         \label{fig:5}
\end{figure}

Figure \ref{fig:4} shows the distribution of our variable $\lambda\equiv\kappa^2\chi M_{\mathrm{ej}}^3/E_{\mathrm{ej}}$ with respect to $\lambda_0$ from our fiducial model, with $0,10^8,10^9$, and $10^{10}$ years passed. Based on this distribution, Fig. \ref{fig:5} shows sample light curves for six values of $\lambda$. 

\section{Observed Rates} \label{sec:observedrates}

We assume that the volumetric rate of stellar collision events takes the form $R(z)=R_0\times f(z)$, where $R_0$ is the rate at redshift zero with units Mpc$^{-3}$ yr$^{-1}$, and $f(z)$ is the redshift evolution. We calculate $R_0$ as the product of the collision rate per galaxy with a SMBH of a certain mass and the volumetric density of galaxies with the same SMBH mass \citep{torrey2015}. We associate each galaxy with a SMBH of mass $M_{\bullet}$ with a halo mass $M_h$ using the following prescription: we calculate the bulge mass associated with $M_{\bullet}$ using Eq. (2) in \cite{mcconnell2013}, the corresponding total stellar mass using Fig. 1 in \cite{bluck2014}, and finally the corresponding halo mass using Eq. (2) in \cite{moster2010}. Given a specific halo mass, we then convert the mass function fit from \cite{warren2006} into a function of redshift $z$, i.e. a function of the form $n(z)=n_0\times f(z)$, where $n$ is the number density of halos at a given mass and $n_0$ is a constant. This method gives us our redshift evolution $f(z)$, completing our calculation of $R(z)$. 

We can then calculate the overall number of events of a given type by integrating over redshift,
\begin{equation}\label{eq:21}
\ N=\int_0^{z}\epsilon\left(z'\right)\frac{4\pi R(z')}{(1+z')}\frac{dV_c}{dz'}dz',
\end{equation}
where $dV_c/dz$ is the comoving volumetric element and $\epsilon(z)$ is the detection efficiency, $0\leq\epsilon(z)\leq1$. $\epsilon(z)$ depends on multiple factors: the survey footprint and cadence, as well as what fraction of detected events can actually be distinguished. For the upcoming Large Synoptic Survey Telescope's (LSST) Deep Drilling Field (DDF) survey, we expect that $\epsilon(z)$ will be no more than $\sim10^{-3}$ at low redshift (and possibly much lower due to the short duration of these events), and will decline monotonically at higher redshift \citep{villar2018}. We note that although much more observing time will be given to the Wide-Fast-Deep (WFD) survey \citep{lsst2019}, we expect that the average revisit time of $\sim3$ days will be too long to identify a significant number of our events, especially at higher energies. 

\begin{figure}
   \includegraphics[width=0.8\hsize]{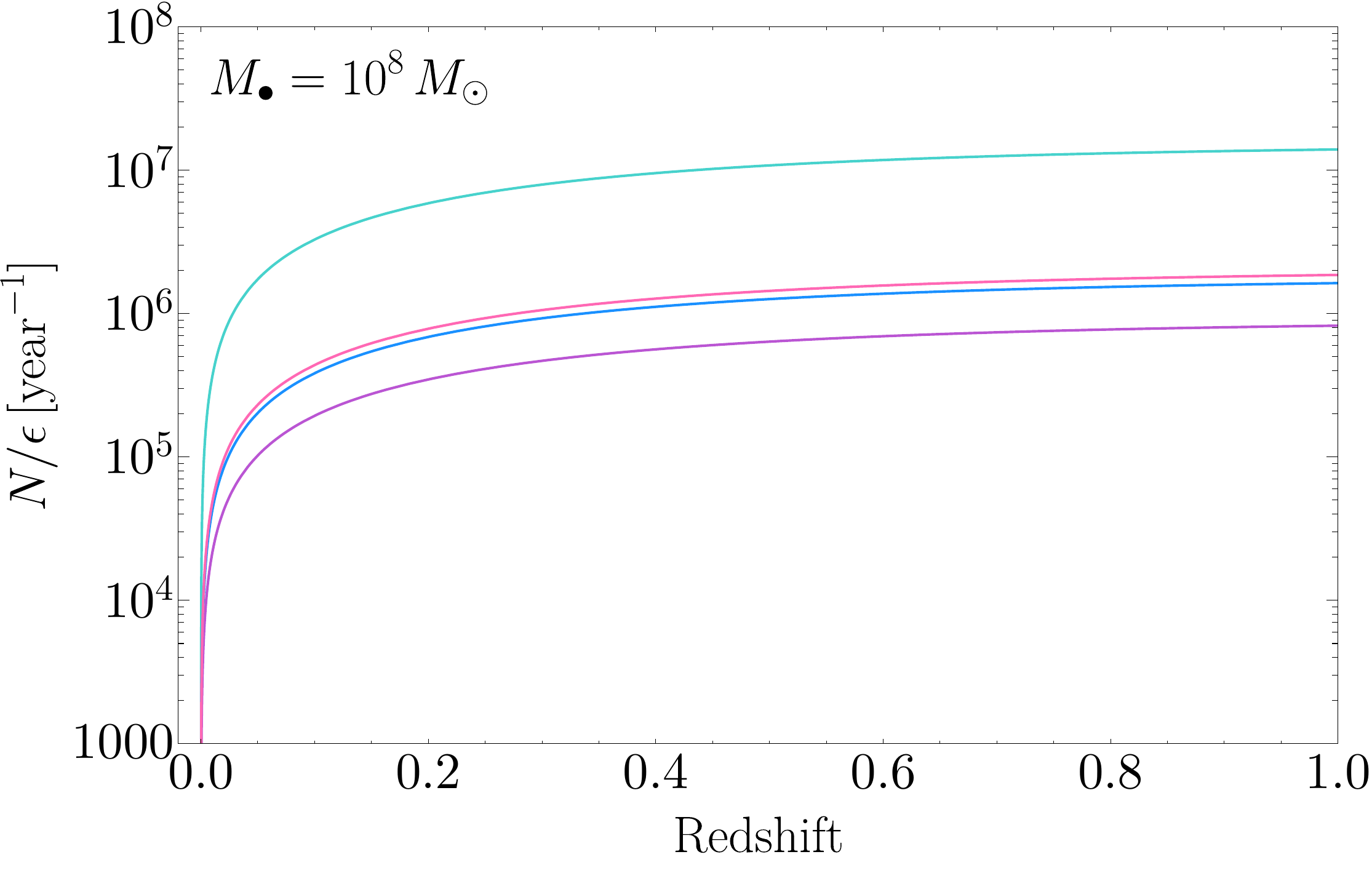}{(a)}
   \includegraphics[width=0.8\hsize]{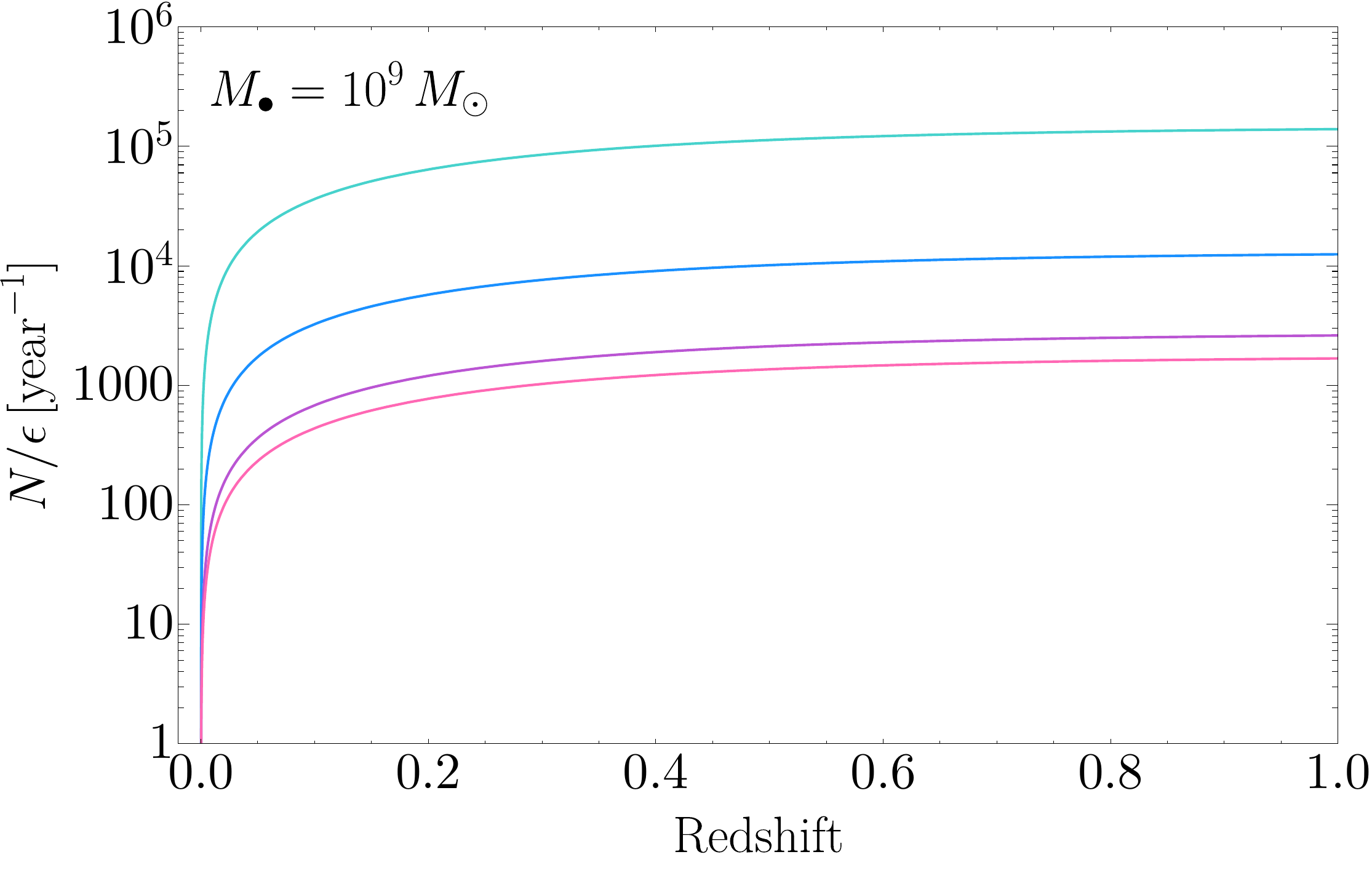}{(b)}
   \includegraphics[width=0.8\hsize]{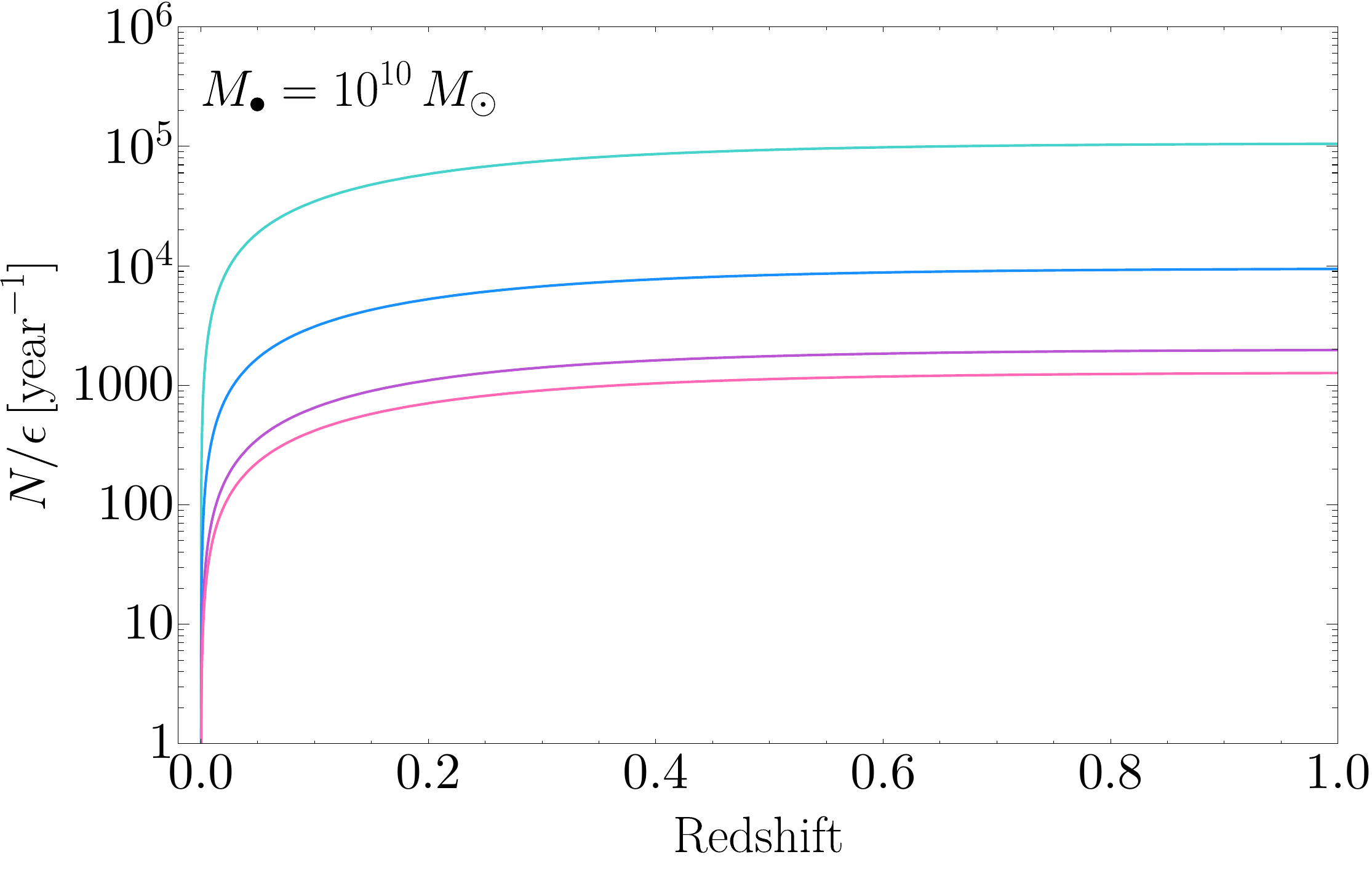}{(c)}
   \includegraphics[width=0.1\hsize]{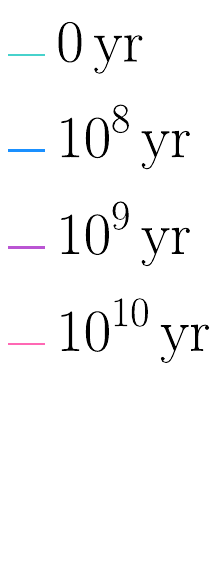}{}
      \caption{The cumulative number of collision events for galaxies with (a) $M_{\bullet}=10^8\,\Msun$, (b) $M_{\bullet}=10^9\,\Msun$, and (c) $M_{\bullet}=10^{10}\,\Msun$, for $0,10^8,10^9$, and $10^{10}$ years passed. We make the simplifying assumption that the detection efficiency $\epsilon(z)$ is a constant in order to move it out of the integral in Eq. \eqref{eq:21} (realistically, for a survey like LSST, we expect it to decline monotonically with redshift).
              }
         \label{fig:6}
\end{figure}

In figure \ref{fig:6}, we integrate over redshift $z$ up to some value and plot $N/\epsilon$ as a function of $z$, making the simplification that $\epsilon(z)$ is a constant. 

\section{Discussion} \label{sec:discussion}

We find that star-star collisions which release $\sim10^{49}-10^{51}$ erg are the most common in the three host galaxies we consider, with $M_{\bullet}=10^8,10^9,10^{10}\,\Msun$. Galaxies with higher-mass SMBHs are more likely to have higher-energy collisions due to the higher velocities near the center of the galaxy, but they have overall lower collision rates due to their lower stellar density. Surveys in the near future could possibly detect several tens of events like these each year \citep{villar2018}. In addition, collisions which release upwards of $\sim10^{53}$ erg can occur with a lower collision rate $\sim10^{-6}$ yr$^{-1}$. These higher-energy collisions would release similar energy as SLSNe \citep{galyam2012}, but with the distinguishing feature of being high-metallicity events due to their occurrence at the center of a galaxy \citep{rich2017}. Conventional SLSNe, on the other hand, are believed to show a preference for low-metallicity environments \citep{leloudas2015, angus2016}. Furthermore, we only expect to find these high-energy, high-velocity stellar collisions in galaxies with a SMBH with mass $M_{\bullet}\gtrsim10^8\,\Msun$, which can be used as a straightforward initial screening for these events. In addition, the most energetic collisions are most likely to take place near the SMBH, which will be an important distinguishing feature when comparing to CCSNe.

For $\lambda/\lambda_0<1$, which we predict represents over half of all possible collisions, the peak luminosity is roughly equal to or even greater than that from most supernovae, but the light curve is expected to decay much faster. At the most extreme values of $\lambda$ among our samples, the light curve could have a peak luminosity roughly equal to that of a SLSNe \citep{galyam2019}, but it would decay over 6 order of magnitude in luminosity in under 2 days, making events like these highly unlikely to be detected. However, some of the most common events we predict, with $\lambda/\lambda_0\sim0.1-1$, could possibly decay slowly enough to be detected. It is possible that they would also be mistaken as low-luminosity supernovae \citep{zampieri2003, pastorello2004}.

Finally, we note that these stellar collisions will likely create a stream of debris that would partly accrete onto the SMBH, creating an accretion flare. This accretion flare may resemble a tidal disruption event (TDE, \cite{loeb1997, gezari2021, dai2021, mockler2021}), even though the black hole is too massive for a TDE. The stellar explosion we have described in this work will be a precursor flare to the black hole accretion flare. We expect that the center of mass of the debris from the stellar collision would follow a trajectory consistent with momentum conservation after the collision and will also spread in its rest frame following the explosion dynamics that we consider. Altogether it would resemble a stream of gas that gets thicker over time. The accretion rate on the SMBH could be super-Eddington as in the case of TDEs and make the black hole shine around or above the Eddington luminosity, $L_E=1.4\times10^{46}\,(M_{\bullet}/10^8\,\Msun)$ erg/s. This luminosity is far larger than we calculated for the collision itself and could be much easier to detect. The details of the accretion flare will be sensitive to the distance of the collision from the SMBH and the velocities and masses upon impact. We leave the numerical and analytical study of this problem to future work (Hu \& Loeb 2024, in prep). In addition, the radiation emitted from these collisions could contribute to the inhabitability of nearby planets, similar to the effect from SMBHs and TDEs \citep{chen2018, forbes2018, pacetti2020}.

\begin{acknowledgements}
      This work was supported by the Black Hole Initiative at Harvard University, which is funded by grants from the John Templeton Foundation and the Gordon and Betty Moore Foundation.
 
\end{acknowledgements}

%
%

\bibliographystyle{aa}
\bibliography{ref}{}

\end{document}